\documentclass[a4paper,11pt]{article}

\usepackage{amsmath,amssymb,amsthm}
\usepackage[dvips]{graphicx}

\begin{document}

\title{\bf 2d manifold-independent spinfoam theory}
\author{{Etera R Livine ${}^{1,2}$, Alejandro Perez ${}^{1,3}$,
Carlo Rovelli ${}^{1,3}$}\\[.2cm]
{\it ${}^{1}$ Centre de Physique Th\'eorique - CNRS, Case 
907,}\\
{\it Luminy, F-13288 Marseille, EU.}\\
{\it${}^{2}$ The Blac6kett Laboratory Imperial College,}\\ 
{\it London SW7 2BZ, EU.}\\
{\it ${}^{3}$ Department of Physics, University of Pittsburgh,}\\ 
{\it Pittsburgh PA 15260, USA.}}
\date{September 28, 2001}
\maketitle

\begin{abstract}
A number of background independent quantizations procedures have
recently been employed in 4d nonperturbative quantum gravity.  We
investigate and illustrate these techniques and their relation in the
context of a simple 2d topological theory.  We discuss canonical
quantization, loop or spin network states, path integral quantization
over a discretization of the manifold, spin foam formulation, as well
as the fully background independent definition of the theory using an
auxiliary field theory on a group manifold.  While several of these
techniques have already been applied to this theory by Witten, the
last one is novel: it allows us to give a precise meaning to the sum
over topologies, and to compute background-independent and, in fact,
``manifold-independent" transition amplitudes.  These transition
amplitudes play the role of Wightman functions of the theory.  They
are physical observable quantities, and the canonical structure of the
theory can be reconstructed from them via a $C^{*}$ algebraic GNS
construction.  We expect an analogous structure to be relevant in 4d
quantum gravity.
\end{abstract}

\section{Introduction}

In this paper we study the quantization of the two-dimensional (2d)
topological field theory characterized by the action
\begin{equation}
    S[\omega,B]=\int Tr[BF], 
    \label{eq:action}
\end{equation}
where $\omega$ is an $SU(2)$ connection field, $F$ its field strength
two-form, and $B$ a scalar field in the adjoint representation of
$SU(2)$.  The theory is the 2d version of BF theory \cite{BF}, and can
be seen as the $e\mapsto 0$ limit ($e$ being the coupling constant) of
2d Yang Mills theory \cite{witten}.  We shall consider this theory on
manifolds of arbitrary topology, without and with boundaries, as well
as in a context which allows us to sum over manifold topologies.

We utilize different quantization strategies, and discuss their
relation.  (For a beautiful introduction to many of the ideas in this
area and a detailed annotated bibliography, see \cite{baez2}.)  First,
we utilize canonical quantization, and exhibit the loop basis
\cite{loop} and the spin network basis \cite{spinnet}.  Next, we
consider a path integral quantization defined in terms of a
triangulation of the 2d manifold, and the spinfoam formalism
\cite{spinfoam}.  We show equivalence with canonical quantization and
we show that this technique allows the theory to be defined on rather
arbitrary manifolds without and with boundaries.  Finally, we recover
the theory from the Feynman expansion of an auxiliary field theory
over a group manifold.

The two dimensional system we consider has been studied in a variety
of ways, and several of the techniques used below are known.  See in
particular \cite{witten}, the book \cite{ambjorn}, and complete
references therein.  The last technique, namely the auxiliary field
theory over a group, on the other hand is quite recent.  This
technique derives from the 2d matrix model approaches to 2d quantum
gravity developed in the context of ``zero dimensional" string theory
\cite{2dg}; the idea was nontrivially extended to 3d by Boulatov
\cite{boulatov} and to 4d by Ooguri \cite{ooguri,ooguri2}, then to the
Barrett Crane model \cite{BC} in \cite{roberto}, and to rather
arbitrary diffeomorphism-invariant but non-topological theories in
\cite{mike}.  The main interest of this technique is that it provides
a natural prescription for summing over topologies.  Using this
technique, we can define and compute the theory's transition
amplitudes in a fully manifold-independent fashion.

These transition amplitudes capture the full physical and gauge
invariant content of the theory.  In this diff-invariant context, they
play a role somewhat analogous to the role that the Wightman functions
play for a quantum field theory over a background.  As for the
Wightman functions \cite{wightman}, one can reconstruct the canonical
structure of the theory from them, using the $C^{*}$ algebraic
Gelfand-Naimark-Segal (GNS) construction.  In addition, we show that
these transition amplitudes are given by the $n$-point functions of
the auxiliary field theory.

A technical result of this paper is the definition and the computation
of these manifold-independent transition amplitudes for the theory
(1).  These turn out to be strictly related to the matrix models ones
-- the precise relation is detailed in the Appendix.  The motivation
for this exercise, on the other hand, is to investigate and illustrate
the different quantization techniques and their relations, in view of
the physical 4d theory.  The auxiliary field theory technique finds
its full motivation in the context of (4d) diff-invariant but
non-topologically invariant theories such as general relativity.  In
this context, it provides the prescription for a sum over
triangulations which restores the triangulation independence of the
transition amplitudes \cite{roberto,mike,alex}.  In the simpler
context studied here, triangulation independence is already guaranteed
by topological invariance, and the auxiliary field theory technique is
used only to define the sum over topologies.  In spite of this
simplification, the study of the 2d context shed considerable light on
several issues emerged in the physically important 4d context,
particularly on the general structure of a diffeomorphism invariant
quantum field theory \cite{JMP,baez}.  We think that the structures
and techniques developed here will play a role in the non-topological,
but still diff-invariant, context of physical quantum gravity.

\section{Classical theory}

Consider the action (1) defined over a given two-dimensional manifold
$\cal M$.  Introducing coordinates $x^a$, $a=1,2$ on $\cal M$ and a
basis of (anti-hermitian) matrices $\tau_{i}$ in $su(2)$, the action
is written in terms of the components of the fields as
\begin{equation}
    S=\int  B ^i F^i_{12}\  dx^{1} dx^2
    \label{eq:actionc}
\end{equation}
(sum over repeated indices is understood.)
The field equations are 
\begin{eqnarray}
dB  &=& 0, \\
F&=& 0,
\end{eqnarray}
where $d$ is the $\omega$-covariant exterior differential. 
Apparently, there are two local gauge invariances in the action: the
conventional Yang-Mills-like local $SU(2)$ transformations generated by
a Lee algebra valued scalar field $\lambda$
\begin{eqnarray}
\delta_{\lambda} \omega = d\lambda, \\
\delta_{\lambda} B  = [B ,\lambda], 
\end{eqnarray}
and the active diffeomorphisms generated by a vector field $v$
\begin{eqnarray}
\delta_{v} \omega = L_{v}\omega, \\
\delta_{v} B  = L_{v}B,  
\end{eqnarray}
$L_{v}$ being the Lie derivative.  However, on shell a diffeomorphism
generated by $v$ is the same transformation as an $SU(2)$
transformation generated by the field
\begin{equation}
    \lambda^i=v^a \omega_{a}^i, 
\end{equation}
as can be easily checked by writing these equations in components and
using the equations of motion.  Therefore in this theory the
diffeomorphisms (acting on the space of solutions) can be considered
as a subgroup of the $SU(2)$ gauge transformations.  Solutions are
given by flat connections and (covariantly) constant $B $ fields. 
Locally, therefore, there is no gauge-invariant degree of freedom.  

\section{Hamiltonian analysis}

Assume now that $\cal M$ has the topology $S_{1}\times R$, and let
$x^a=(t,\phi)$, where $t$ is a non-compact coordinate along $R$ and
$\phi\in[0,2\pi]$ is a periodic coordinate on $S_{1}$.  We can then
write the action (\ref{eq:actionc}) as
\begin{equation}
    S=\int dt \int d\phi\  (B ^i \partial_{t} \omega^i_{\phi} -
    \omega^i_{t} DB ^i),
\end{equation}
where $D$ is the covariant derivative of the connection
$A^i=\omega^i_{\phi}$ on the 1d manifold $S_{1}$.  From this we read
out the canonical structure of the theory: the canonical fields are
$A^i(\phi)$ and $B ^i(\phi)$, with Poisson brackets
$\{A^i(\phi),B ^j(\phi')\}=\delta^{ij}\delta(\phi,\phi')$, and there is
the single first class constraint which generates (fixed time) 
$SU(2)$ gauge transformations 
\begin{equation}
    C^{i}= DB^i.
    \label{eq:constraint}
\end{equation}
This confirms that the $SU(2)$ transformations exhaust the gauge
invariances of the theory.  The extended phase space on which this
constraint is defined is the infinite dimensional space of the initial
data $(A^i(\phi),B ^i(\phi))$. The space of the gauge orbits is 
finite dimensional. In fact, it is two dimensional. It can be 
coordinatized by the two gauge invariant quantities 
\begin{eqnarray}
T &=& Tr U[A] = Tr\  P \, e^{\oint_{S_{1}}\!\! A}, \\
L &=& B ^i(0)B ^i(0), 
\end{eqnarray}
where $0$ is an arbitrary point, since $B ^i(\phi)B ^i(\phi)$ is
constant in $\phi$.  The two quantities $T$ and $L$ commute with the
constraint and form a complete system of gauge invariant observables. 
The physical states of the theory are therefore characterized by these
two quantities.  All relevant information follows from this.  For
instance: given two set of initial data $(A^i_{in}(\phi),
B_{in}^i(\phi))$ and $(A^i_{out}(\phi), B_{out}^i(\phi))$, can the
first evolve into the second?  The answer is yes if and only if
$T[A^i_{in},B_{in}]=T[A^i_{out},B_{out}]$ and $L[A^i_{in},B
_{in}]=L[A^i_{out},B_{out}]$, and in this case the trajectory between
the two in the coordinate time $t$ is given by any arbitrary one
parameter family of gauge transformations, parametrized by $t$, taking
the $in$ fields into the $out$ fields.

\section{Canonical quantization}

We begin to construct the quantum theory by quantizing the
unconstrained hamiltonian theory and imposing the quantum constraint
\`a la Dirac.  The space of the unconstrained quantum states is formed
by functionals of the connection $\Psi[A]$.  The $B $ field is
represented by the functional derivative operator $B (\phi)=-i\hbar
\delta/\delta A(\phi)$, which gives the right commutation relations. 
The constraint is the generator of gauge transformations on the
argument of the quantum state, and therefore the states that solve the
Dirac constraint are the states of the form
\begin{equation}
    \Psi[A]=f(Tr U[A])=\psi(U[A]). 
\end{equation}
Therefore a physical state is given by a function $\psi$ over $SU(2)$,
depending only on the trace of its argument -- that is, invariant under
the adjoint action of the group over itself: $\psi(U)=\psi(VUV^{-1})$. 
Such functions are denoted ``class functions".  There is a natural
invariant scalar product on these states, which is
\begin{equation} 
    \langle\Psi,\Psi'\rangle=\int dU\ \overline{\psi(U)}\ \psi'(U). 
\end{equation} 
where $dU$ is the Haar measure on $SU(2)$.  Throughout the paper, we
denote $\cal H$ the Hilbert space of the square integrable functions
over $SU(2)$.  Then the physical Hilbert space is
\begin{equation} 
   {\cal H}_{ph} = {{\cal H}\over SU(2)}={L_{2}[SU(2)]\over
   SU(2)},
   \label{hilbert}
\end{equation} 
where the $SU(2)$ in the denominator is the adjoint action of the
group over itself.  The gauge invariant observables $T$ and $L$ are
well defined on $\cal H$.  The first gives
\begin{equation}
    T \psi(U)= Tr(U)\  \psi(U); 
\end{equation}
and a straightforward calculation shows that the second gives 
\begin{equation}
    L \psi(U)= \hbar^2 C \psi(U),
\end{equation}
where $C$ is the $SU(2)$ Casimir operator.  The essential condition on
the physical scalar product on the solution of the constraint equation
is that the physical operators be self-adjoint.  They are, and
therefore the scalar product we have defined is the physically correct
one.  Since the Casimir of $SU(2)$ has eigenvalues $j(j+1)$, with half
integer $j$, we obtain immediately a first result from the quantum
theory: the observable $L$ is quantized, with eigenvalues
\begin{equation}
    L_{j} = \hbar^2\ j(j+1). 
\end{equation}
A natural basis can be obtained by diagonalizing $L$.  Since the
Casimir operator is diagonal over the irreducible representations, we
define
\begin{equation} 
    \psi_{j}(U)=Tr_jU \equiv Tr\,[R^{(j)}(U)],
\end{equation}
where $R^{(j)}(U)$ is the matrix representing $U$ in the representation
$j$. Using the well known relation 
\begin{equation} 
    \int dU\ \overline{R^{(j)}_{ab}(U)}\ R^{(k)}_{cd}(U)=\frac{1}{2j+1} \ 
    \delta^{jk}\ \delta_{ac}\delta_{bd},
    \label{orto}
\end{equation}
we have immediately that the $\psi_{j}$ form an orthonormal basis,
which we denote $|j\rangle$. Therefore
\begin{equation}
\langle U|j\rangle= Tr_{j}U
\label{eq:Uj}
\end{equation} 
We have thus $\langle j|j'\rangle = \delta_{jj'}$.  The action of $T$
on this basis can be obtained directly from standard Clebsch Gordon
technology.  The normalized (generalized) eigenstates $|T\rangle$ of
the $T$ operator form a continuos orthonormal basis, and thus we have
\begin{equation} 
    \langle T'|T\rangle = \delta(T,T').
    \label{ortoT}
\end{equation}
A useful set of operators to consider is the one corresponding to the 
classical observables 
\begin{equation} 
    T_{j} = Tr_jU[A]. 
\end{equation}
These are clearly diagonal in the $U$ representation and
\begin{equation} 
    T_{j}|0\rangle = |j\rangle. 
\end{equation}

Although this model is very simple, it has remarkable analogies with
the the loop quantization of general relativity.  In loop quantum
gravity as well, indeed, we have states functionals of the connection
that can be written as functions of the holonomies $U$ of the
connection.  While in loop quantum gravity one must consider
holonomies along different loops, here there is essentially only one
loop: the one that wraps around $S_{1}$.  Thus, this theory can be
seen as a sort of ``single loop" loop quantum gravity.  Indeed,
(\ref{eq:Uj}) is an elementary loop transform.  Better, the basis
$|j\rangle$ that we have introduced is precisely the ``spin network
basis" \cite{spinnet}, in this simplified case of a single loop.  The
$T$ operator is then analogous to the quantum gravity loop operator,
and the operator $L$ is analogous to the area operator, which is also
given in terms of the Casimir of $SU(2)$, and which is diagonalized by
the spin network basis \cite{area}.  The quantization of $L$ is thus
the 2d analog of the quantization of the area.

\section{Discretization}

Let us now consider a completely different path for quantization.  We
start with the covariant theory, triangulate the manifold ${\cal M}$ and
consider a lattice gauge theory that discretizes our theory.  In
general, the discretization introduced in going to the lattice kills
the degrees of freedom below a certain scale; however, since the
theory we are considering has no local degrees of freedom, we expect
nothing to be really lost in the discretization.  As we shall see,
this will indeed turn out to be the case.  In addition, the
discretization allows us to get rid of the restriction to the simple
$S_1 \times R$ topology, and consider arbitrary topologies with an
arbitrary number of boundaries.

Let us therefore fix a triangulation $\Delta$ of $\cal M$.  A
triangulation in two dimensions is formed by triangles, edges, and
points.  A triangle is bounded by three edges and an edge by two (not
necessarily distinct) points.  An edge bounds precisely two triangles
and a point bounds an arbitrary number of edges.  It is more
convenient to use the dual $\Delta^*$ of a triangulation $\Delta$,
which is formed by trivalent vertices, links and faces (or plaquettes)
with an arbitrary number of sides.  We discretize the connection by
replacing it with a group element $U_{l}$ for every link $l$ of
$\Delta^*$.  We discretize the field $B$ by replacing it with a Lie
algebra valued variable $B_{f}$ for every face $f$.  Finally, we
approximate the action (\ref{eq:action}) as a sum over the faces 
\begin{equation} 
    S[B_{f}U_{l}] = \sum_{f} Tr [B_{f}U_{f}], 
    \label{eq:discreteaction}
\end{equation}
where $U_{f}=U_{f_{1}}\ldots U_{f_{n}}$ when $f_{1}\ldots {f_{n}}$ are
the links around the face $f$.  To first order in the area of the
plaquettes we have $Tr [B_{f}U_{f}] \to Tr[B(1+F)]=Tr[BF]$.  We then
consider the quantum theory defined by the partition function
\begin{equation}
    Z_{\cal M}=\int dB_{f}\ dU_{l}\ e^{Ñ{i\over \hbar}S[B_{f}U_{l}]}. 
\label{pf}
\end{equation}
The subscript ${\cal M}$ indicates the manifold: we will indeed check
shortly that $Z_{\cal M}$ depends on the manifold but not the
triangulation.  The integral over $B_{f}$ can be performed explicitly
\cite{alexint}, giving
\begin{equation}
    Z_{\cal M} =\int dU_{l}\ \prod_{f} \delta(U_f) =\int dU_{l}\
    \prod_{f} \delta(U_{f_{1}}\ldots U_{f_{n}}),
\label{pla}
\end{equation}
where $\hbar$ and $2\pi$ coefficients have been absorbed in a
redefinition of the measure $dB_{f}$.  We can then expand the delta
function over $SU(2)$, using a well known representation of it
\begin{equation}
    Z_{\cal M} =\int dU_{l}\ \prod_{f} \sum_{j} (2j+1)\  Tr_j U_{f}. 
\label{j}
\end{equation}
 Equation (\ref{j}) can be rearranged as 
\begin{equation}
    Z_{\cal M} = \sum_{j_{f}} \int dU_{l}\ \prod_{f} (2j_{f}+1)\
    Tr_{j_{f}} U_{f}.
\label{coloring}
\end{equation}
where $j_{f}$ is an assignment of a spin $j$ to every face $f$, and
the sum is over all such assignments.

These two steps have replaced the integrals over the continuous
variables $B_{f}$ (one per face) with sums over the discrete index
$j_{f}$ (one per face).  To understand how this may have happened,
consider the following analogy.  Let $-\pi<x<\pi$ and consider a space
$S$ of sufficiently regular functions over this interval.  Consider
the integral
\begin{equation}
\delta(x) = \frac{1}{2\pi} \int_{-\infty}^{\infty} dp\ e^{ipx}.
\end{equation}
It converges to a well defined distribution over $L$: the delta 
function on the origin. However, on this restricted interval the same 
delta function can be obtained with a discrete sum as 
\begin{equation}
\delta(x)  = \frac{1}{2\pi} \sum_{n} e^{inx}.
\end{equation}
In a sense, the values $p=n$ of $p$ are ``sufficient" for the
integral.  Similarly, the ``quantized" values $j$ in (\ref{coloring})
are sufficient to give the same delta function over the group as the
one defined by the $B$ integral in (\ref{pf}).

The remaining integrals in (\ref{coloring}) can be performed because
each link $l$ always bounds precisely two faces, and therefore each
integration variable $U_{l}$ enters in precisely two traces.  Using
(\ref{orto}), we conclude that the representation must be the same for
every two adjacent faces, which is to say for all faces.  Therefore
the sum over ${j_{f}}$ reduces to a single sum over $j$.  From
equation (\ref{orto}), each integral (that is, each link) contributes
a factor $1/(2j+1)$, and a bunch of delta functions that end up
contracted among themselves.  A moment of reflection on the pattern of
the indices in the integrals will convince the reader that after all
the integrals have been performed, there remains one trace for each
(trivalent) vertex in $\Delta^*$.  Each such trace gives a
contribution $(2j+1)$ to the integral.  Putting all together we obtain
\begin{equation}
    Z_{\cal M} = \sum_{j} 
 \prod_{f} (2j+1)
 \prod_{l} 1/(2j+1)
 \prod_{v} (2j+1) = 
 \sum_{j} (2j+1)^{F-L+V}, 
\end{equation}
where $F,L,V$ are the numbers of faces links and vertices in
$\Delta^*$.  The quantity $\chi=F-L+V$ is a topological invariant,
that is, for a fixed manifold is triangulation independent.  In fact,
it is the Euler characteristic $\chi({\cal M})$ of $\cal M$.  For a
compact oriented surface of genus $g$, the Euler characteristic is
$\chi=2-2g$.  Thus
\begin{equation}
    Z_{\cal M} = \sum_{j} (2j+1)^{\chi({\cal M})} = \sum_{j}
    (2j+1)^{2-2g({\cal M})},
    \label{genus}
\end{equation}
which converges for $g>1$, namely for all Riemann manifolds except for
the sphere ($g$=0) and the torus ($g=1$).
For non-orientable surfaces,
the Euler characteristic and the genus
are linked by $\chi=2-g$
and the partition function converges for $g>2$, namely for
all surfaces but the projective plane $RP^2$ ($g=1,\chi=1$)
and Klein's bottle $K$ ($g=2,\chi=0$).
As we will see in a moment,
this divergence is harmless and physical quantities are all well
defined.

\section{Boundaries}

The calculation in the previous section is not very meaningful by
itself, since the sourceless partition function $Z_{\cal M} $ is either
to be normalized to one or infinite.  We have performed it only as a
preliminary step to compute something more interesting.  To get to
something more interesting we have to have states in the theory and
transition amplitudes.
\begin{figure}[ht]
    \centering
%{\psfig{figure=uno.eps,height=4cm}}
\includegraphics[height=4cm]{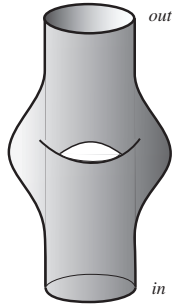}
\caption{A spacetime manifold of genus 1 with two boundary 
    components.}
    \label{fig:1}
\end{figure}

A diffeomorphism invariant field theory has the following general
structure.  Let the manifold $\cal M$ have boundary $\Sigma$, formed
by $n$ connected components $\Sigma_{i}$ with $i=1,\ldots,n$.  Figure
1 illustrates, as an example, a manifold with genus one and a two
component boundary.  (Unconstrained) initial data $A_{i}$ can be
associated to each connected component $\Sigma_{i}$ of the boundary. 
The path integral of the bulk variables at fixed boundary values gives
the transition amplitudes $Z_{n}[A_{i}]$.  To make the notation
lighter we do not explicitly indicate the manifold dependence of these
quantities, but it is important to remember that they do depend on the
topology of the spacetime manifold that interpolates between the
boundaries (unlikely the transition function of the next sections,
which do not).  In particular, $Z_{1}[A]$ defines the
``Hartle-Hawking" vacuum state state \cite{hh}, and
$Z_{2}[A_{out},A_{in}]$ defines the propagator.  This propagator
defines the projection operators $P$ that projects the space of the
unconstrained states $\psi(A)$ into the physical Hilbert space,
\begin{equation}
    P\psi(A) = \int dA'\ Z_{2}[A,A'] \psi(A') 
    \label{eq:P}
\end{equation}
and defines the  physical scalar product of the theory 
\begin{equation}
    \langle \psi | \psi' \rangle = \int dA\ dA'\   
    \overline{\psi(A)} Z_{2}[A,A'] \psi'(A') . 
    \label{eq:sp}
\end{equation}
Therefore $Z_{1}[A]$ picks a preferred state in the Hilbert space,
$Z_{2}[A_{out},A_{in}]$ maps the Hilbert space associated to the $in$
boundary to the Hilbert space associated to the $out$ boundary,
$Z_{3}[A_{1},A_{2},A_{3}]$ gives a three legs transition amplitude,
and so on.  

The amplitudes $Z_{n}[A_{i}]$ are related to each other as follows. 
Assume the two manifolds ${\cal M}_{1}$ and ${\cal M}_{2}$ can be
glued along a common boundary $\Sigma_{j}$ obtaining a manifold ${\cal
M}$.  The transition amplitudes of ${\cal M}$ can be obtained from the
ones of ${\cal M}_{1}$ and ${\cal M}_{2}$ by integrating on the common
initial data $A_{j}$ on $\Sigma_{j}$.

More formally, this relation can be expressed as follows \cite{atiyah}. 
There is a category $M$ whose objects are collections of circles, and
whose maps are 2d manifolds having these circles as boundaries.  The
diffeomorphism invariant field theory is then a representation of this
category, that is, a functor from $M$ to the category $H$ of the
Hilbert spaces (whose maps are linear maps between Hilbert spaces).

In the case we are considering, all boundary connected components are
isomorphic isomorphic to $S_{1}$, and therefore there is a single
fundamental Hilbert $\cal H$ space in the game (The Hilbert space
associated to $n$ circles is the symmetric tensor product of $n$
copies of $\cal H$).  Also, using the discretized definition of the
path integral, all integrals are well defined.  The dual triangulation
$\Delta^*$ of $\cal M$ induces a triangulation of each boundary
component.  We can always assume that this triangulation is made by a
single segment, and denote the associated group element as $U_{i}$. 
Boundary values are therefore group elements $U_{i}$, which we can
immediately identify as the holonomy of the connection around the
boundary.  All integration measures are given by the Haar measure. 
Let us compute the amplitudes $Z_{n}[U_{i}]$.

The Hartle-Hawking amplitude $Z_{1}[U]$ of a hemisphere can be
computed immediately from (\ref{pla}).  Inserting the boundary value
and discretizing the manifold with a single face, we have immediately
\begin{equation}
    Z_{1}[U] = \sum_{j} (2j+1)\ Tr_{j}U = \delta(U). 
\end{equation}
Thus, the Hartle Hawking state $|HH\rangle$ is the delta function on
the group.

Next, consider a manifold with the topology of a cylinder.  Again, we
can discretize the cylinder with a single face, bounded by the
two boundaries of the manifold and by an internal link joining the two
boundaries.  In (\ref{pla}), we have then a single integral and a
single trace
\begin{eqnarray}
Z_{2}[U_{out},U_{in}]&=& \int dV\
 \delta(U_{in}^{-1}VU_{out}V^{-1})\   \nonumber \\ &=&
\sum_{j} (2j+1)\int dU Tr[R^{(j)}(U^{-1}_{in}\ U U_{out}\ U^{-1})].
\end{eqnarray}
Using (\ref{orto}) again, this gives 
\begin{equation}
   Z_{2}[U_{out},U_{in}]
= \sum_{j}\  Tr_jU_{out}\  Tr_jU_{in}.  
\end{equation}
This is precisely the projector on the class functions over the group 
\begin{eqnarray}
\int dU' \    Z_{2}[U,U']\ \psi(U)
 &=&  
\int dU'dV\  \delta(U^{-1}VU'V^{-1})\ \psi'(U')
\nonumber \\ &=&
\int dV\  \psi'(VUV^{-1}). 
\end{eqnarray}
If $\psi$ and $\psi'$ are class functions
\begin{equation}
\langle \psi | \psi' \rangle =  
    \int dU\ \overline{\psi(U)}\ \psi'(U). 
\end{equation}
That is, we recover (\ref{hilbert}), the same physical Hilbert space
${\cal H}_{ph}$ as in the canonical theory. 

In the basis $|j\rangle$
\begin{equation}
    Z_{n}(j_{i})\equiv 
    \int dU_{i}\  Z_{n}(U_{i}) \  Tr[R^{(j)}(U_{i})], 
    \label{eq:Zj}
\end{equation}
the Hartle Hawking state is 
\begin{equation}
    Z_{1}(j) = \psi_{HH}(j)= \langle j | HH \rangle  = 2j+1,
\end{equation}
and the propagator 
\begin{equation}
    Z_{2}(j,j') = \langle j | j' \rangle = \delta_{j\, j'}.
\end{equation}

The higher transition amplitudes, as well as the transition amplitudes
for manifolds with an arbitrary number of holes, can be computed in a
similar fashion using the discretization of the manifold.  A shortcut
can e taken using the functorial properties of the transition
amplitudes.  For instance, let us glue a hemisphere to a cylinder,
obtaining a hemisphere.  Correspondingly, we should obtain a Hartle
Hawking state by propagating a Hartle Hawking state
$\psi_{HH}(j)=\sum_{j}Z_{2}(j,j')\psi_{HH}(j)$.  In coordinate space 
\begin{equation}
    \int dU'\ Z_{2}(U,U')\ Z_{1}(U') =  Z_{1}(U);
\end{equation}
explicitly
\begin{equation}
    \int dV dU'\ \delta(UVU'V^{-1})\ \delta(U') =  \delta(U). 
\end{equation}
Let us now cut out a disk from a cylinder.  We obtain a
manifold with three boundaries.  The corresponding amplitude
$Z_{3}(j,j',j")$ must satisfy
\begin{equation}
Z_{2}(j,j')= \sum_{j"} Z_{3}(j,j',j") Z_{1}(j"), 
\end{equation}
and we have immediately 
\begin{equation}
Z_{3}(j,j',j") = \frac{1}{2j+1}\  \delta_{j,j',j"}
\end{equation}
where $\delta_{j,j',j"}$ is one if $j=j'=j"$, and zero otherwise. 
And, similarly, we obtain 
\begin{equation}
Z_{n}(j_{i}) = \frac{1}{(2j+1)^{n-2}}\  \delta_{j_{i}}
\end{equation}
where $\delta_{j_i}$ is one if all $j_{i}$ are the same and zero
otherwise.
In term of boundary group elements,  one gets:
\begin{equation}
Z_n(U_i)=\int dV_i\ \delta\left(\prod_{i=1}^nV_i^{-1}U_iV_i\right)
=\int dV_2..dV_n\ \delta(U_1V_2^{-1}U_2V_2..V_n^{-1}U_nV_n)
\end{equation},
that is the delta function on the product of the conjugacy classes
of the boundary holonomies.

\medskip

Then we can deduce the higher genus case by gluing some punctures together.
For example, considering $Z_3$ and identifying two punctures together,
one gets the partition function for the punctured torus $Z^{(g=1)}_1$,
and so on.

This way, we get the partition function for
an orientable (compact) surface of genus $g$
\begin{equation}
Z^{(g)}=\int \prod_{a=1}^g dC_adD_a \
\delta\left(\prod_i C_aD_aC_a^{-1}D_a^{-1}\right),
\end{equation}
where  the group elements $C_i,D_i$ are associated to the $2g$
non-trivial loops/cycles of the surface.
Then one immediately recovers equation (\ref{genus})
$$
Z^{(g)}=\sum_j(2j+1)^{2-2g}.
$$
Let us point out that taking the cylinder or two-punctured sphere,
we can get both the torus or Klein's bottle by gluing the two ends together,\
which implies that the two surfaces have the same partition function.
Indeed they have the Euler characteristic $\chi=0$.

Similarly, we can start with a manifold of genus $g$ and cut out a 
disk. By the gluing technique, we immediately get:
\begin{equation}
    Z^{(g)}_{1}(U)=\int \prod_{a=1}^g dC_adD_a \ \int dV\
\delta\left(VUV^{-1}\prod_a C_aD_aC_a^{-1}D_a^{-1}\right),
\end{equation} 
which leads to
\begin{equation}
    Z^{(g)}_{1}(j)= (2j+1)^{1-2g}, 
\end{equation} 
And we finally get easily the general expression for the amplitude of
a genus $g$ manifold with $n$ disk removed, and therefore $n$ boundary 
components
\begin{equation}
   Z^{(g)}_{n}(U_i)=\int \prod_{a=1}^g dC_adD_a \
\int \prod_{i=1}^n dV_i\
\delta\left(\prod_i V_iU_iV_i^{-1}\prod_a C_aD_aC_a^{-1}D_a^{-1}\right)
\end{equation}
\begin{equation}
    Z^{(g)}_{n}(j_{i})= (2j+1)^{2-2g-n} \ \delta_{j_{i}} 
\end{equation}
Notice that all these transition functions are finite. 

The above discussion illustrates in detail the relation with canonical
quantization.  In particular, notice that the propagator associated to
the cylinder is precisely the projector on the solutions of the
quantum constraint.  Notice that this is precisely the structure in
the formal functional quantization of general relativity in Hawking's
approach.  Notice also that, as it is always the case in diffeomorphism
invariant theories,\\
--- the matrix elements of the physical scalar product,\\
--- the projector on the physical states and \\
--- the evolution operator in coordinate time \\
are all identified.

The difference between a topological field theory and a field theory
which is diffeomorphism invariant, but is not topological, is only in
the number of the degrees of freedom involved.  In other words, the
general structure that we expect is the same, but the Hilbert space
associated to a boundary has a much richer structure.  Furthermore, if
we define the theory by means of a discretization, we do not expect
triangulation independence, because a fixed triangulation cuts off the
number of degrees of freedom.

The quantization we have discussed reproduces, in simplified form, the
state-sum definition of $BF$ theory in 3d and 4d, as in the models
developed by Ponzano and Regge and by Turaev and Viro, and by Turaev,
Ooguri, Crane and Yetter \cite{PRTV}.  The analog for general
relativity is given by the state sum formulations of Barret and Crane
\cite{BC} and their variants \cite{alex}, where, however, one looses
triangulation invariance.  Notice in particular the sum over
assignments of representations to faces in equation
(\ref{coloring}).  This sum corresponds to the state sum of these
models.  Here, of course, the ``Clebsch Gordon" condition on the links
are simplified by the fact that only two faces join at a link and
therefore the representation of the two faces must be same, thus
collapsing the sum over arbitrary colorings to a sum over a single
$j$. 

\section{Auxiliary field theory}

We now come to a main part of the paper.  Following the ideas in
\cite{boulatov,ooguri,roberto,mike} we define an auxiliary field
theory on a group manifold, whose Feynman graph expansion gives 
the above 2d $BF$ theory.

Let $g_{i}$, for $i=1,2$ be in $SU(2)$; we change notation for the
group elements for consistency with the conventions in this area and
also to emphasize the different role that group elements assume now. 
Consider a real scalar field $\Phi(g_{1},g_{2})$ on $SU(2)\times
SU(2)$, having the following two properties.  Symmetry
\begin{equation}
    \Phi(g_{1},g_{2}) = \Phi(g_{2},g_{1}),
    \label{eq:symmetry}
\end{equation}
and right $SU(2)$ invariance
\begin{equation}
    \Phi(g_{1},g_{2}) = \Phi(g_{1}g,g_{2}g), \ \ \ \ \forall g\in 
    SU(2).
    \label{eq:invariance}
\end{equation}
These two symmetries can also be expressed by writing the field as
\begin{equation} 
    \Phi(g_{1},g_{2}) = \int_{SU(2)} dg \ 
    \left(\Psi(g_{1}g,g_{2}g)+\Psi(g_{2}g,g_{1}g)\right),
    \label{eq:intinvariance}
\end{equation}
where $\Psi(g_{1},g_{2})$ is an arbitrary field on $SU(2)\times
SU(2)$.  The field theory is defined by the nonlocal action
\begin{eqnarray} 
S &=& \int_{SU(2)\times SU(2)} dg_{1}dg_{2}\ \Phi^2(g_{1},g_{2}) 
\nonumber \\ &&
+\frac{\lambda}{3!} \int_{SU(2)\times SU(2)\times SU(2)}
dg_{1}dg_{2}dg_{3}\ \Phi(g_{1},g_{2}) \Phi(g_{2},g_{3})
\Phi(g_{3},g_{1}).
    \label{eq:actft}
\end{eqnarray}
Notice that in each of the two terms each integration variable is
shared by precisely two fields.  The structure of the two terms,
``kinetic" and ``potential", can be represented as in Figure 2, where
the circles represent the fields and the lines represent their shared
arguments which are integrated over.
\begin{figure}[ht]
    \centering
%{\psfig{figure=F2.eps,height=3cm}}
\includegraphics[height=3cm]{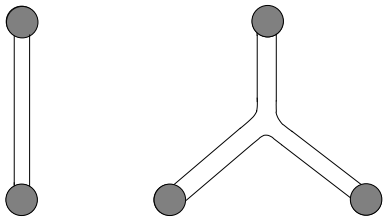}
\caption{Structure of kinetic and potential term in the action.}
    \label{fig:2}
\end{figure}

We now study the perturbative Feynman expansion of the theory. 
Because of the symmetries of the field, care should be taken in
inverting the kinetic term on the subspace of the symmetric fields
only.  On this subspace the kinetic term is indeed diagonal, and it is
therefore straightforward to see that the propagator $P$ and the
vertex $V$ can be written as
\begin{equation}
    P(g_{1},g_{2};g'_{1},g'_{2})=\int dg \left(\delta(g_{1}g,g'_{1})
    \delta(g_{2}g,g'_{2})  + \delta(g_{1}g,g'_{2})
    \delta(g_{2}g,g'_{1})\right) 
    \label{eq:prop}
\end{equation}
and
\begin{equation}
    V(g_{1},g_{2};g'_{1},g'_{2};g"_{1},g"_{2})= 
    \delta(g_{1},g'_{2})
    \delta(g'_{1},g"_{2})
    \delta(g"_{1},g_{2}) 
    \label{eq:vertex}
\end{equation}
By representing a delta function with a line, with the two end points
representing the two arguments, and the group integration as two dots,
we can represent the propagator and the vertex as in Figure 3.
\begin{figure}[ht]
    \centering
%{\psfig{figure=F3.eps,height=2cm}} 
\includegraphics[height=2cm]{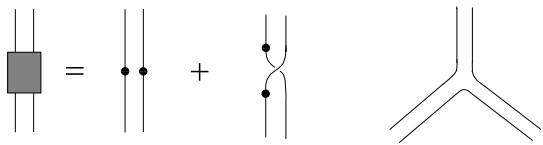}
\caption{Propagator and vertex.}
    \label{fig:3}
\end{figure}

Consider the Feynman graph expansion of the the partition function
\begin{equation}
    Z=e^{F} =\int d\Phi\ e^{-S[\Phi]}, 
    \label{eq:Za}
\end{equation}
The partition function $Z$ is the sum of the amplitudes of all closed
Feynman graphs.  The ``Free energy" $F$ is the given by the connected
graphs.  Consider a connected Feynman graph of order $V$, that is with
$V$ vertices.  It will have $E=\frac{3}{2}V$ propagators.  Because of
the sum in (\ref{eq:prop}), which symmetrizes the arguments of the two
deltas, the amplitude of the graph is the sum of $2^E$ terms.  Each of
these terms can be represented by replacing the propagator in Figure 3
with one or the other of the terms in the sum, in each of the $E$
propagators.  Consider one of these terms, as for example in Figure 4.
\begin{figure}[ht]
    \centering
%{\psfig{figure=F4.eps,height=3cm}}
\includegraphics[height=3cm]{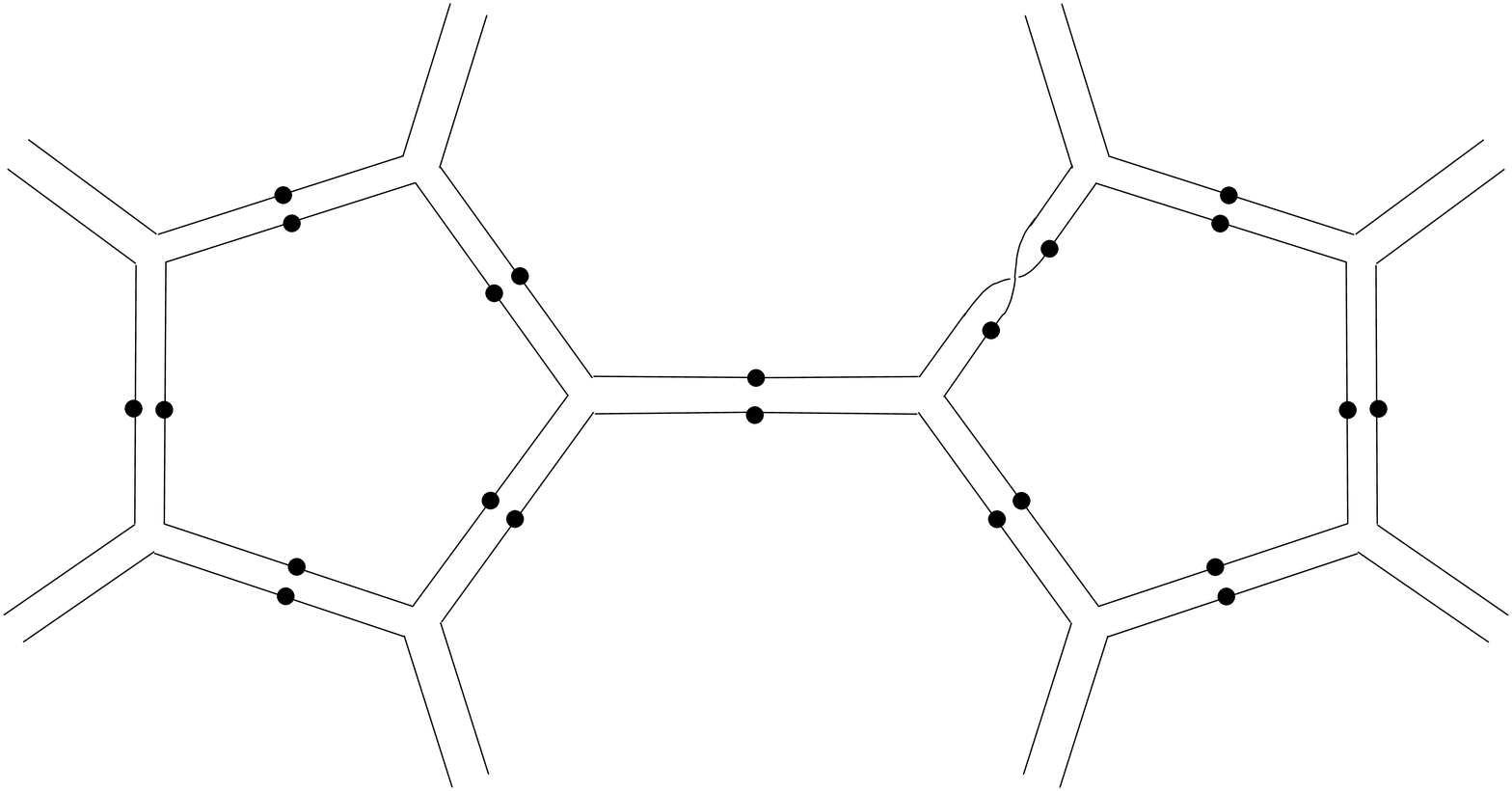}
\caption{A 2d two-complex (part of it).}
    \label{fig:4}
\end{figure}
Notice that the corresponding graphical representation is formed by a
graph on which closed lines run along propagators and through
vertices.  We denote ``face" one of these closed lines, and say that
it is bounded by the ``edges" formed by the propagators along which
the line runs.  The collection $\Gamma$ of faces, links and vertices
with their boundary relations, defines a two-complex, characterized by
the fact that each vertex is three-valent (it bounds three edges) and
each edge is bi-valent (it bounds two faces).  We denote a two complex
with trivalent vertices and bivalent edges as a 2d two-complex.  The
Feynman amplitude of the 2d two-complex $\Gamma$ is just given by a
multiple integral of $2E+3V$ delta functions.  Each integration
variable enters in two delta functions.  The pattern in which these
are enchained is simply given by the closed lines around the faces. 
In other words, there is a closed sequence of deltas for each face. 
By integrating away all the variables at the end points of the
propagators, we have then easily that the amplitude of the two-complex
is
\begin{equation}
    A_\Gamma = \int dg_{e}\ \ \prod_{f}\ \delta(g_{f1} \ldots g_{fn})
    \label{eq:amplitude2c}
\end{equation}
where ${f1}, \ldots, {fn}$ are the edges that bound the face $f$. 

Now, consider a 2d manifold $\cal M$ with a triangulation $\Delta$. 
Notice that the dual triangulation $\Delta^{*}$ is precisely a 2d
two-complex $\Gamma$.  Furthermore, note that if $\Gamma$ is the dual
of a triangulation of $\cal M$, then the amplitude
(\ref{eq:amplitude2c}) is {\em precisely\/} the partition function of
BF theory on $\cal M$, given above in equation (\ref{pla}). That is
\begin{equation}
    Z_{\cal M}=  A_{\Delta^*}
    \label{eq:main}
\end{equation}
if $\Delta$ is a triangulation of $\cal M$.  This is a key result. 

The dual of a triangulation of a two-manifold is a 2d two complex.  In
two dimensions, the converse is true as well.  Namely each 2d
two-complex is the dual of the triangulation of a 2d manifold
\cite{roberto2}.  The same is not true in higher dimensions, where one
is lead to consider arbitrary two-complexes as generalized manifolds,
but such complications are absent in 2d.

Just to get a feeling of the relation between Feynman graphs and 2d
manifolds, consider the example of a simple graph formed by two
vertices connected by three propagators.  In expanding the
symmetrization in the propagators, we obtain $2^3$ terms.  Let us
analyze first the one in which none of the lines crosses in a planar
representation, as in Figure 5a.  It is immediate to see that this is
the dual of a simple triangulation of a sphere obtained by gluing two
triangles along their perimeter.  Indeed, the two-complex has three
faces, and the triangulation has therefore three (bivalent) vertices. 
The Euler number is $\chi(\Gamma)=F-E+V=3-3+2=2$ and the genus is
zero.  Next, consider the term in which there is a crossing in each
propagator, as in Figure 5b.  Following the line, we see that this two
complex has a simple face: the corresponding triangulation has a
single vertex, and a moment of reflection shows that it triangulates a
torus.  Indeed, the Euler number is $\chi(\Gamma)=1-3+2=0$ and the
genus is one.  Finally, consider the case in which only one propagator
has a crossing (Figure 5c).  In this case there are two faces.  The
corresponding manifold is $RP^2$, the non orientable manifold obtained
from a disk by identifying opposite points on the perimeter.  The
Euler number is $\chi(\Gamma)=2-3+2=1$.
\begin{figure}[ht]
    \centering
%{\psfig{figure=F5.eps,height=3cm}} 
\includegraphics[height=3cm]{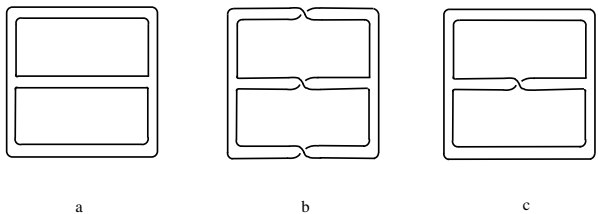}
\caption{Two-complexes corresponding in 2d to the sphere $S$, the torus $T$, and 
the projective plane $RP^2$} 
    \label{fig:5}
\end{figure}

We can then immediately perform the integrals in (\ref{eq:amplitude2c})
as we did in the previous sections, obtaining
\begin{equation}
    A_{\Gamma} = \lambda^V \  (2j+1)^{\chi(\Gamma)}
    \label{eq:amplitude}
\end{equation}

Now, let $N(V,\chi)$ be the number of 2d two-complexes with Euler
number $\chi$ obtained by expanding all graphs with $V$ vertices. 
This number is clearly finite and its determination is, in principle,
a well defined combinatorial problem.  The sum $Z$ over the connected
Feynman diagrams gives
\begin{equation}
    Z =  \sum_{\chi} N_\chi \ \sum_{j} (2j+1)^\chi,
    \label{eq:amplitudeft}
\end{equation} 
where the weight factor $N_\chi$ is given by 
\begin{equation}
    N_{\chi} =  \sum_{V}\ N(V,\chi)\  \lambda^V.  
\end{equation}
This defines a version of the topological field theory in which a sum
over the manifold topologies is naturally implemented.  The resulting
theory does not depend on any fixed underlying manifold structure. 
The different spacetime manifold topologies are generated as Feynman
graphs of the auxiliary theory.  The auxiliary field theory fixes a
well defined prescription for the sum over these topologies.  The
possibility of transitions through disconnected spacetime manifolds
(for instance, two circles going into two circles via two cylinders)
is taken into account by the standard field combinatorics of the
connected/disconnected (reducible/irreducible) graphs.  In the
appendix, we discuss the precise relation between this theory and the
matrix models.  As shown in the appendix, the transition amplitudes
between states of color $j$ are essentially the transition amplitudes
of a matrix model with $N=2j+1$ dimensional matrices
\cite{ambjorn,2dg,ooguri2}.  For the discussion of their finiteness,
see for instance \cite{ambjorn,ooguri2}.

\section{Transition amplitudes}

Once more, the physical content of the theory is not in its sourceless
partition function but in its transition amplitudes, which can be
obtained by inserting field operators in the path integral, coupling a
source and taking derivatives with respect to the source, or computing
the partition functions with fixed boundary conditions on the fields
on spacetime boundaries.  Remarkably, the transition amplitudes of the
topological theory, obtained by defining the theory on a manifold with
boundaries, are directly connected to the $n$-point functions of the
auxiliary theory, obtained by inserting fields in the path integral.

The easiest way to see this, is to fix the gauge in the auxiliary
field.  Notice indeed that because of the invariance
(\ref{eq:intinvariance}), we can consider the quantity
\begin{equation}
    \psi(g)=\Phi(1,g),
    \label{eq:psiuno}
\end{equation}
or, equivalently
\begin{equation}
    \psi(g_{2}g_{1}^{-1})=\Phi(g_{1},g_{2}),
    \label{eq:psidue}
\end{equation}
which has the same information as the field. We thus define the 
$n$-point functions of the auxiliary field theory as 
\begin{equation}
    W(g^{(1)},\ldots,g^{(n)})=Z^{-1}\ 
    \int \ [D\Phi]\ \psi(g^{(1)}) \ldots \psi(g^{(n)})\
    e^{-S[\Phi]}.
   \label{eq:npoint}
\end{equation} 
The momentum space version of the transition functions are
\begin{equation}
     W_{j_1 \ldots j_n} =Z^{-1}\  \int [D\Phi]
\ \Phi_{j_{1}}\ldots \Phi_{j_{n}}
     \ \ e^{-S[\Phi]},
    \label{eq:final}
\end{equation}
where 
\begin{equation}
     \Phi_{j} = \int d{g_{1}}d{g_{2}}\ Tr_{j}(g_{2}g^{-1}_{1})\ 
     \Phi(g_{1},g_{2}). 
\end{equation}
One could consider the more general observables
\begin{equation}
     \Phi_{j}^{(\alpha)} = \int d{g_{1}}d{g_{2}}\
Tr_{j}\left((g_{2}g^{-1}_{1})^\alpha\right)\ 
     \Phi(g_{1},g_{2}),
\end{equation}
however $Tr_j(g^\alpha)$ is simply a linear combinaison
of traces $Tr_k(g)$ with $k$ ranging from 0 to $\alpha j$
(due to the decomposition into irreducible representations
of the representation $j\otimes j \otimes..\otimes j$).

Consider an irreducible Feynman graph in the perturbative expansion of
$W(g^{(1)},\ldots,g^{(n)})$.  A moment of reflection, will convince
the reader that this is given by the same graphs as before, with the
difference that now $n$ propagator lines are open.  The group element
$g^i$ is associated to the $i$-th open propagator line, and in
expanding the enchainment of the deltas, the two individual lines of
the propagator get connected to each other through $g^i$.  This can be
represented by a circle formed by a {\em single\/} line, colored with
$g^i$.  Now, this is precisely the representation of a dual
triangulation of a manifold with $n$ boundaries, colored with group
elements $g^1, \ldots, g^n$.  Therefore the Feynman expansion of the
$n$-points function is the sum of the transition amplitudes of the
topological theory, computed on manifolds with $n$ boundaries, and
summed over all topologies of the interpolating spacetime manifold.

Going to momentum space, we have then immediately, for the 
irreducible $n$-point function $\Gamma_{j_1\ldots j_n}$
\begin{equation}
    \Gamma_{j_1\ldots j_n} = \sum_{\chi} N_{\chi,n}\ (2j+1)^{\chi-n}
\ \delta(j_{i})
    \label{eq:np}
\end{equation}
where
\begin{equation}
    N_{\chi,n}= \sum_{m} N(m,\chi,n) \lambda^m, 
\end{equation}
where $N(m,\chi,n)$ is the number of irreducible (that is, connected)
2d two-complexes with $n$ boundaries, $m$ vertices and Euler number
$\chi$.  The relation between the transition amplitudes and their
irreducible part can be obtained by standard methods: we define the
generating functional of the connected graphs as a function of a 
source class function $J(g)$ with components $J_{j}$. 
\begin{equation}
    \Gamma[J] = \sum_{n}\sum_{j_1\ldots j_n}\ \Gamma_{j_1\ldots j_n} 
    J^{j_{1}}\ldots J^{j_{n}}
    \label{eq:gencon}
\end{equation}
and the generating functional of the transition amplitudes
\begin{equation}
    W[J] = \sum_{n}\sum_{j_1\ldots j_n}\  W_{j_1\ldots j_n} 
    J^{j_{1}}\ldots J^{j_{n}}.
    \label{eq:gen}
\end{equation}
And we have 
\begin{equation}
    W[J]=e^{\Gamma[J]}= Z^{-1}\ 
    \int \ d\Phi\ e^{-S[\Phi]+J\cdot\Phi},
    \label{eq:generating}
\end{equation}
where 
\begin{equation}
  J\cdot\Phi=\sum_{j}\ \Phi_{j}J^{j}= \int dg_{1}dg_{2}\
  \Phi(g_{1},g_{2})\ J(g_{2}g^{-1}_{1}).
    \label{eq:source}
\end{equation}
Notice that the divergence due to the sphere and the torus in $Z$ does
not affect the transition functions, since it is always divided out by
the $Z^{-1}$ factor in $(\ref{eq:npoint})$; equivalently, closed
disconnected Feynman diagrams do not contribute to the transition 
amplitudes.

We have thus obtained the explicit form of all the transition
amplitudes, in a form that is independent from the topology of the
underlying spacetime manifold.  Rather, these can be viewed as the
transition amplitudes computed by summing over all topologies of the
interpolating spacetime manifold.  The prescription for the relative
weights is implicitly fixed by the auxiliary field theory.  These
transition amplitudes can be directly computed as the $n$-point
functions of the auxiliary field theory.  The sum defining transition
amplitudes can be viewed as a sum over the colored 2d two-complexes. 

A $n$-dimensional analog of (\ref{eq:actft}) is obtained by taking a
field which is a function of $n$ group elements, and a potential term
of order $\frac{1}{2}n(n-1)$ having the structure (see Figure 2) of
an $n$ dimensional simplex.  The expansion of the Feynman graphs
generates then $n$d two-complexes.  The dual of an $n$ dimensional
triangulation is an $n$d two-complex.  This construction provides a
manifold independent definition on $BF$ theory in $n$ dimensions. 

What is particularly remarkable is that the constraint that reduced 4d
$BF$ theory to general relativity can be obtained in the auxiliary
field theory simply by requiring an additional invariance of the field
under a subgroup for the field over the group \cite{roberto,alex}. 
The 4d sum is therefore over colored 4d two-complexes.  These are
complexes in which each edge bounds four faces and each vertex bounds
ten edges.  The representations associated to the faces can vary
freely provided that Clebsch-Gordon conditions are respected at the
edges.  Edges carry the additional degree of freedom given by the
intertwiner between the representations associated to the adjacent
faces.  Such colored two complexes are denoted spinfoams and
transition amplitudes can therefore be expressed as sums over
spinfoams, hence the denomination spinfoam models.

The remarkable aspect of this formalism, when applied to gravitational
theories, is that a spinfoam admits an interpretation as a discretized
4-geometry, and therefore the sum over spin foams turns out to be a
well defined version of the Misner-Hawking sum of geometries
\cite{Misner-Hawking}.  In particular, it turns out that one of the
Casimirs of the representation associated to the face represents the
area of the face, thus giving a metric interpretation to the coloring. 
This can happen because, as we have seen the Casimir is quadratic in
the $B$ field.  The constraint that reduces $BF$ theory to general
relativity forces the $B$ field to be the product of two tetrad
fields, and thus the Casimir is a product of four tetrads associated
to a face, and a straightforward calculation shows that it is the
square of the area of the face.  Furthermore, the boundary Hilbert
space turns out to be precisely the Hilbert space of loop quantum
gravity, thus providing a precise link with the canonical formalism. 
The hope is that the auxiliary field theory technique could allow us
to define and compute 3-geometry to 3-geometry transition amplitudes
as sums over spinfoams, where the sum takes into account the full sum
over topologies.  As we have seen this hope is realized at least in
the very simple context of the 2d topological theory.

\section{Canonical and algebraic structure}

Since the spacetime boundary can be composed by an arbitrary number of
circles, a generic state of the system is the symmetric tensor product
of an arbitrary number of copies of $\cal H$.  That is, it is the Fock
space $\cal F$ over $\cal H$.
\begin{equation}
    {\cal F} = \bigoplus_{n=0,\infty} \left({\cal H}_{1}\otimes_{s}
    \ldots \otimes_{s} {\cal H}_{n}\right),
    \label{eq:Fock}
\end{equation}
where all ${\cal H}_{i}$'s are isomorphic to $\cal H$.  A basis in
$\cal F$ is given by the vectors $|j_{1}\ldots j_{n} \rangle$, where
$n$ is arbitrary and the set is ordered.  The vacuum state $|O
\rangle$ (not to be confused with the $|j=0\rangle$ state in $\cal
H$), that is, a normalized vector of the $n=0$ term in
(\ref{eq:Fock}), represents the absence of any boundary.

What are the physical operators on this Hilbert space and to which
physical observables do they correspond?  Recall that we identified
only two observables in the canonical analysis of the classical
theory, $T$ and $L$.  However, that result was under the assumption
that the physical state is defined on a boundary formed by a single
component.  If the boundary has $n$ components, there must be $n$
different values $T_{1},\ldots, T_{n}$ and $L_{1},\ldots, L_{n}$ of
these observables, representing the trace of the holonomy of the
connection and the length of the (constant) $B$ field in each of the
components.  In other words, the classical phase space becomes the
disjoint collection of an infinite set of components, labelled by the
number of circles $n$.  In each of these, the observables can be taken
to be $T_{1},\ldots, T_{n}$ and $L_{1},\ldots, L_{n}$.  It is also
convenient to define an observable $N$ that takes the value $n$ on the
$n$-th phase space component.  The operator $N$ and all the $T_{i}$'s
are simultaneously diagonalized, by the basis $|j_{1}\ldots j_{n}
\rangle$.

We now introduce the additional operator $\mathrm{T}_{j}$, which
increases $n$ by one:
\begin{equation}
    \mathrm{T}_{j}|j_{1}\ldots j_{n} \rangle = |j_{1}\ldots j_{n},j
    \rangle.
\end{equation}
Intuitively, this operator can be understood as follows.  When
measuring the holonomy on a boundary with $n$ circles, we have to
specify on which circle we are measuring it.  There are therefore $n$
distinct holonomy operators acting on $|j_{1}\ldots j_{n} \rangle$,
each acting on a different $j_{i}$.  But this is not sufficient, since
we can also measure the holonomy of a next extra circle opening up,
and $\mathrm{T}_{j}$ is related to this operation.

Observable quantities of the theory are the transition functions
$W_{j_1 \ldots j_n}$.  These have a straightforward physical
interpretation as follows.  We can arbitrarily divide the $n$ indices
$j_1,\ldots, j_n$ into two families: the $in$ ones $|j^{in}_{1}\ldots
j^{in}_{n_{in}} \rangle$ and the $out$ ones $|j^{out}_{1}\ldots
j^{out}_{n_{out}} \rangle$.  Then $W_{j_1 \ldots j_n}$ is the
probability amplitude to find the system in the $|j^{out}_{1}\ldots
j^{out}_{n_{out}} \rangle$ state after having found it in the
$|j^{in}_{1}\ldots j^{in}_{n_{in}} \rangle$.  

Notice that we can write 
\begin{equation}
    |j_{1}\ldots j_{n} \rangle = T_{j_{1}}\ldots T_{j_{n}}
    |O \rangle
    \label{eq:TO}
\end{equation}
and 
\begin{equation}
    W_{j_1 \ldots j_n} = 
    \langle 0 | T_{j_{1}}\ldots T_{j_{n}}|O\rangle
    \label{eq:0TO}
\end{equation}
The $W_{j_1 \ldots j_n}$ functions, and their coordinate
space transform $W(g^{(1)},\ldots,g^{(n)})$, are therefore the vacuum
expectation values of products of $T_{j}$ operators.

Conversely, assume that the $W(g^{(1)},\ldots,g^{(n)})$, functions are
given to us.  Then we can reconstruct the quantum theory from them, in
the spirit of Wightman.  To do that, consider a linear $\cal L$ space of
sufficiently regular ``test" functions $f(g^{(1)},\ldots,g^{(n)})$.  We
can promote $\cal L$ to a $C^*$ algebra by defining the adjoint
$f^*(g^{(1)},\ldots,g^{(n)})=\overline{f(g^{(n)},\ldots,g^{(1)})}$,
the norm $|f|=$ $ sup[f(g^{(1)},\ldots,g^{(n)})]$, the product $(fh) 
(g^{(1)},\ldots, g^{(n+m)}) = f(g^{(1)},\ldots,g^{(n)})$ $  
h(g^{(n+1)},\ldots,g^{(n+m)})$.  The $W(g^{(1)},\ldots,g^{(n)})$
functions define a positive linear functional $W$ on $\cal L$ by
\begin{equation}
    W(f)= \sum_{n} \int 
    dg^{(1)}\ldots dg^{(n)}\ 
    W(g^{(1)},\ldots,g^{(n)})
    f(g^{(1)},\ldots,g^{(n)})
\end{equation}
We can thus run the GNS construction and obtain an Hilbert space, an
algebra of operators and a vacuum state such that $W$ is the vacuum
expectation value of the operators in the algebra.  We can reconstruct
in this way the hamiltonian structure defined above.  Therefore the
hamiltonian structure of the theory can be reconstructed from the
transition amplitudes, in the spirit of Wightman.  

Notice that the pre-Hilbert scalar product $(f,h)=W(f^*h)$ that the
GNS construction is degenerate, because of the $g\mapsto g'gg'{}^{-1}$
invariance under conjugation of the two point function and therefore
$\cal L$ is projected to ${\cal L}_{ph}$, which is formed by class
functions only.  For instance, the function $f(g)=f^{ab}R_{ab}(g)$,
with $Tr f=0$, has zero norm and is projected out of $L_{ph}$ in the
GNS construction of the Hilbert space.  ${\cal L}_{ph}$ is then
spanned by the basis $f^{j_{1}\ldots j_{n}}$ of the components of the
class functions $f$, and
\begin{equation}
    W(f)= W_{j_1 \ldots j_n}f^{j_{1}\ldots j_{n}}.
    \label{eq:Wf}
\end{equation}

In conclusion, the $W_{j_1 \ldots j_n}$ functions are gauge invariant,
capture the full content of the theory and correspond to physical
observables with a clear interpretation: they are therefore a natural
set of objects in terms of which to deal with the theory and its
physical content.  Analogous objects for 4d quantum gravity exist and
will be described elsewhere.

\section{Conclusions and perspectives}

We have studied the quantization of a simple diffeomorphism invariant
theory using several techniques, which turn out to be nicely
consistent.  The auxiliary field theory method provides a prescription
for a manifold-independent definitions of the theory, and a way to
compute manifold-independent transition amplitudes.  These can be
understood as sums over different spacetime topologies.  They capture
the physical content of the theory, and represent physical observable
quantities \cite{observables}.  

The interest of this auxiliary field theory technique is that it
extends to quantum gravity.  Indeed, there is an auxiliary field
theory formulations of quantum gravity \cite{roberto,mike,alex}.  In
this case, the Feynman graph sum provides a genuine sum over
triangulations, which erases the triangulation dependence of the
non-topological theory.  Furthermore, the Hilbert space ${\cal
H}_{ph}$ associated to a boundary component turns out to be precisely
the Hilbert space of loop quantum gravity.  In particular, the $W$
functions of quantum gravity capture its full diff-invariant physical
content, and we expect that they could be obtained as the $n$-point
functions of an auxiliary field theory such as the one defined in
\cite{alex}.  Thus, a beautifully overall covariant and canonical
coherent picture of nonperturbative quantum gravity seems to be
emerging.

A word of caution should be added, regarding the general use of the
auxiliary field theory and the possibility of computing transition
amplitudes in this path integral form.  As we have seen, the two point
function is the projector on the solutions of the canonical
constraints.  The projector should have zero eigenvalues.  These
correspond to the unphysical gauge degrees of freedom, which should be
projected out.  As well known, gauge degrees of freedom may give
divergences in the path integral, essentially due to the integration
over the infinite volume of the gauge group.  These may be cured; in
the theory in this paper, care has been taken in inverting the
propagator on the physical subspace only, thus getting rid of an
infinity.  In a more complicated case, getting rid of those infinities
might not be as simple and zeros of the projector might appear as
divergences.  This issue might be harder to deal with if the potential
term alters the invariances of the kinetic term.

Finally, it would be important to study the auxiliary field theory
formulation and the algebraic structure in the case of an arbitrary
group in which irreducible representations are not conjugate to
themselves.  This might affect the past/future structure of the
transition amplitudes, and the star relation in the $C^{*}$ algebra.

\vskip.4cm

\centerline{------------------------Ñ}

\vskip.4cm

This work was partially supported by NSF Grant PHY-9900791 and by the
Andrew Mellon predoctoral felloship.

\section*{Appendix: Relation with the matrix models}

Let us expand the field in mode components, using the Peter-Weyl 
theorem, as follows 
\begin{equation}
    \Phi(g_1,g_2)=\phi^{a_1b_1a_2b_2}_{j_1j_2}
    \ \overline{R^{j_1}_{a_1b_1}(g_1)}
    \ R^{j_2}_{a_2b_2}(g_2). 
    \label{eq:deco}
\end{equation}
The symmetry property (\ref{eq:invariance}) of the field implies
\begin{eqnarray}
    {\Phi}(g_1,g_2)&=&\int dg\ \phi(g_1g,\ g_2g)
    \noindent \\ &=&
    \phi^{a_1b_1a_2b_2}_{j_1j_2}\  \overline{R^{j_1}_{a_1c_1}(g_1)}
    R^{j_2}_{a_2c_i}(g_2)
    \int dg\  \overline{R^{j_1}_{c_1b_1}(g)}\ R^{j_2}_{c_2b_2}(g).
\end{eqnarray}
Using (\ref{orto}), we can write
\begin{equation}
    {\Phi}(g_1,g_2)=\sqrt{2j+1}\ 
    \phi_j^{a_1a_2}\ \overline{R^{j}_{a_1c}(g)}\ R^{j}_{a_2c}(g)
\end{equation}
where we have defined 
\begin{equation}
    \phi_j^{a_1a_2} = \frac{1}{\sqrt{2j+1}}\  
    \phi^{a_1b_1a_2b_2}_{jj_2}\ \delta_{b_{1}b_{2}}\
    \delta^{jj_{2}}.
\end{equation}
The symmetry property (\ref{eq:symmetry}) imples $\phi_j^{a_1a_2}=
\overline{\phi_j^{a_2a_1}}$. Writing the action (\ref{eq:actft}) in
terms of these modes, we obtain for the kinetic term
\begin{equation}
\int {\Phi}^2(g_1,g_2) dg_1 dg_2 = \phi_j^{a_1a_2}\phi_j^{a_1a_2},  
\end{equation}
and for the potential term 
\begin{equation}
    \frac{\lambda}{3!}\int {\phi}(g_1,g_2){\phi}(g_2,g_3)
    {\phi}(g_3,g_1) dg_1dg_2dg_3=
    \frac{\lambda}{3!} \frac{1}{\sqrt{2j+1}}
    \phi_j^{ab}\phi_j^{bc}\phi_j^{ca}
\end{equation}
The action is a sum over $j$ of terms $S_{j}$.  For each of these
terms, we define an hermitian matrix $M$, of dimension $N=2j+1$ by
$M_{ab}=\phi_j^{ab}$.  Then each term takes the form
\begin{equation}
    S_{j}=\frac{1}{2}{Tr}(M^2)
    +\frac{\lambda}{3!}\frac{1}{\sqrt{N}}{Tr}(M^3)
\end{equation} 
which is a standard form for the matrix models action
\cite{2dg,ooguri}.  Therefore our theory is formed by a collection of
non interacting matrix models, one per representation.

When calculating the Feynman diagrams from the above model, the
propagators (corresponding to the edges of the dual triangulations)
don't give any weight, the vertices give each a weight $N^{-1/2}$ and
the loops in the diagrams (corresponding to faces of the dual
triangulation) give the weight $N$ (size of the matrices), so that the
total weight of a diagrams is 
\begin{equation}
    {\cal W}=\lambda^{V}N^{F-\frac{1}{2}V}
\end{equation}
Using $2E=3V$, we find that the exponent is $F-\frac{1}{2}V=F-E+V$,
the Euler characteristic of the manifold triangulated by the (dual of
the) Feynman diagram.  Notice also that, if one scales the variables
$\phi_j^{ab}$ or $M_a^b$ by a factor $\alpha$, the propagators gets a
factor $\alpha^{-2}$ and the vertices a factor $\alpha^3$, so the
overall factor will be $\alpha^{3V-2E}=1$ and does not affect the
total weight.

\end{document}